\documentstyle[12pt]{article}
\input amssym.def
\input amssym.tex

\begin{document}
\newcommand{\ux}{\!\begin{array}[t]{l}\!x\\\widetilde{ }
\end{array}\!}
\newcommand{\uxk}{\!\begin{array}[t]{l}\!x_k\\\widetilde{ }
\end{array}\!}
\newcommand{\uxm}{\!\begin{array}[t]{l}\!x_{k-1}\\
\widetilde{ }\end{array}\!}
\newcommand{\uxp}{\!\begin{array}[t]{l}\!x_{k+1}\\
\widetilde{ }\end{array}\!}

\newcommand{\wx}{\widetilde{x}}

\begin{titlepage}{\LARGE
\begin{center} A discrete time relativistic Toda lattice
 \end{center}} \vspace{1.5cm}
\begin{flushleft}{\large Yuri B. SURIS}\end{flushleft} \vspace{1.0cm}
Centre for Complex Systems and Visualization, Unversity of Bremen,\\
Postfach 330 440, 28334 Bremen, Germany\\
e-mail: suris \@ mathematik.uni-bremen.de\\  \\ \\ \\
{\small {\bf Abstract.} Four integrable symplectic maps approximating two
Hamiltonian flows from the relativistic Toda hierarchy are introduced.
They are demostrated to belong to the same hierarchy and to examplify
the general scheme for symplectic maps on groups equiped with quadratic
Poisson brackets. The initial value problem for the difference
equations is solved in terms of a factorization problem in a group.
Interpolating Hamiltonian flows are found for all the maps.}

\end{titlepage}

\setcounter{equation}{0}
\section{Introduction}
Although the subject of integrable symplectic maps received in the recent
years a considerable attention, the order in this area seems still to lack.
Given an integrable system of ordinary differential equations with such
attributes as Lax pair, $r$--matrix and so on, one would like to construct
its difference approximation, desirably also with a (discrete--time analog of)
Lax pair, $r$--matrix etc. Recent years brought us several successful
examples of such a construction [1--8], but still not the general rools
and recipes, not to say about algorithms.

Recently there appeared for the first time examples where the Lax matrix of the
discrete--time approximation {\it coincides} with the Lax matrix of the
continuous--time system, so that the discrete--time system belongs to the
{\it same} integrable hierarchy as the underlying continuous--time one
(systems of Calogero--Moser type [7,8]). We want to present here
one more example of this type, which can be studied in full (and beautiful)
details, -- the discrete--time analog of the relativistic Toda lattice [9],
see also [10--12].

The paper is organized as follows. In sect. 2,3 we recall some facts about
the continuous--time relativistic Toda lattice, its $r$--matrix structure
and the solution in terms of a factorization problem in a matrix group.
The most part of these facts is by now well known, but it turned out to
be rather difficult or even impossible to find them in literature in the
form suitable for our present purposes. In sect. 4 we introduce the equations
of motion of the discrete--time relativistic Toda lattice and discuss their
symplectic structure. Sect. 5 contains the Lax pair representation for our
system, and in sect. 6 we give the solution of the initial value problem for
our
system in terms of a factorization problem in a matrix group.

\setcounter{equation}{0}
\section{Relativistic Toda lattice}
The relativistic Toda lattice with the coupling constant $g^2\in{\Bbb R}$
is described by the Newtonian equations of motion
\begin{equation}
\ddot{x}_k=
\dot{x}_{k+1}\dot{x}_k\frac{g^2\exp(x_{k+1}-x_k)}{1+g^2\exp(x_{k+1}-x_k)}-
\dot{x}_k\dot{x}_{k-1}\frac{g^2\exp(x_k-x_{k-1})}{1+g^2\exp(x_k-x_{k-1})},
\quad 1\le k\le N.
\end{equation}
with one of the two types of boundary conditions: open--end,
\[
x_0\equiv\infty, \quad x_{N+1}\equiv-\infty,
\]
or periodic,
\[
x_0\equiv x_N, \quad x_{N+1}\equiv x_1.
\]
It is a known fact (although usually not stressed in the literature) that the
equation (2.1) may be put into the Hamiltonian form in two different ways,
 which lead to two different Hamiltonian functions belonging, remarkably,
to one and the same integrable hierarchy.

The first way to introduce the variables $p_k$ canonically conjugated to
$x_k$ is:
\begin{equation}
\exp(p_k)=\frac{\dot{x}_k}{1+g^2\exp(x_{k+1}-x_k)},
\end{equation}
which leads to the system
\begin{eqnarray*}
\dot{x}_k & = & \exp(p_k)(1+g^2\exp(x_{k+1}-x_k)), \\
\dot{p}_k & = & g^2\exp(x_{k+1}-x_k+p_k)-g^2\exp(x_k-x_{k-1}+p_{k-1}),
\end{eqnarray*}
a Hamiltonian system with the Hamiltonian function
\begin{equation}
J_+=\sum_{k=1}^N \exp(p_k)(1+g^2\exp(x_{k+1}-x_k)).
\end{equation}
The second way to introduce the momenta $p_k$ is:
\begin{equation}
\exp(p_k)=-\frac{1+g^2\exp(x_k-x_{k-1})}{\dot{x}_k},
\end{equation}
which leads to the system
\begin{eqnarray*}
\dot{x}_k & = & -\exp(-p_k)(1+g^2\exp(x_k-x_{k-1})), \\
\dot{p}_k & = & -g^2\exp(x_{k+1}-x_k-p_{k+1})+g^2\exp(x_k-x_{k-1}-p_k),
\end{eqnarray*}
a Hamiltonian system with the Hamiltonian function
\begin{equation}
J_-=\sum_{k=1}^N \exp(-p_k)(1+g^2\exp(x_k-x_{k-1})).
\end{equation}

The Lax representation and the integrability for the flows with the
Hamiltonians (2.3),(2.5) are dealt with in the following statement.
Introduce two $N$ by $N$ matrices
depending on the phase space coordinates $x_k, p_k$ and
(in the periodic case) on the additional parameter $\lambda$:
\begin{eqnarray*}
L & = & \sum_{k=1}^N \exp(p_k)E_{kk}+\lambda\sum_{k=1}^N E_{k+1,k},\\
U & = & \sum_{k=1}^N E_{kk}-\lambda^{-1}\sum_{k=1}^N
g^2\exp(x_{k+1}-x_k+p_k)E_{k,k+1}.
\end{eqnarray*}
Here $E_{jk}$ stands for the matrix whose only nonzero entry on the
intersection
of the $j$th row and the $k$th column is equal to 1. In the periodic case we
set $E_{N+1,N}=E_{1,N}, E_{N,N+1}=E_{N,1}$; in the open--end case we set
$\lambda=1$, and $E_{N+1,N}=E_{N,N+1}=0$. Consider also following two matrices:
\begin{equation}
T_+=LU^{-1}, \quad T_-=U^{-1}L.
\end{equation}

{\bf Theorem 1.} {\it The flow with the Hamiltonian (2.3) is equivalent to
the following matrix differential equations:
\[
\dot{L}=LB-AL, \quad \dot{U}=UB-AU,
\]
which imply also
\[
\dot{T}_+=\left[ T_+,A\right], \quad \dot{T}_-=\left[ T_-,B\right] ,
\]
where
\begin{eqnarray*}
A & = & \sum_{k=1}^N(\exp(p_k)+g^2\exp(x_k-x_{k-1}+p_{k-1}))E_{kk}+\lambda
\sum_{k=1}^N E_{k+1,k},\\
B & = & \sum_{k=1}^N(\exp(p_k)+g^2\exp(x_{k+1}-x_k+p_k))E_{kk}+\lambda
\sum_{k=1}^N E_{k+1,k}.
\end{eqnarray*}
The flow with the Hamiltonian (2.5) is equivalent to
the following matrix differential equations:
\[
\dot{L}=LD-CL, \quad \dot{U}=UD-CU,
\]
which imply also
\[
\dot{T}_+=\left[ T_+,C\right], \quad \dot{T}_-=\left[ T_-,D\right] ,
\]
where
\begin{eqnarray*}
C & = & -\lambda^{-1}\sum_{k=1}^N
g^2\exp(x_{k+1}-x_k-p_{k+1}+p_k)E_{k,k+1}, \\
D & = & -\lambda^{-1}\sum_{k=1}^N g^2\exp(x_{k+1}-x_k)E_{k,k+1}.
\end{eqnarray*}
}

So we see that either of the matrices $T_{\pm}$ (they are in fact connected
by means of a similarity transformation) serves as the Lax matrix for
{\it both} the flows (2.3), (2.5). Note also that the Hamiltonians $J_{\pm}$
belong to the set of invariant functions of $T_{\pm}$, as it is easy to
check that
\[
J_+={\rm tr}(T_{\pm}), \quad J_-={\rm tr}(T_{\pm}^{-1}).
\]

It is often convenient to use instead of the canonically conjugated
variables $x_k, p_k$ another set of variables $c_k, d_k$ defined as
\begin{equation}
d_k=\exp(p_k), \quad c_k=g^2\exp(x_{k+1}-x_k+p_k),
\end{equation}
which satisfy the Poisson brackets
\[
\{c_k,c_{k+1}\}=-c_kc_{k+1}, \quad \{c_k,d_{k+1}\}=-c_kd_{k+1}, \quad
\{c_k,d_k\}=c_kd_k
\]
(only the non-vanishing brackets are written down). In terms of these
variables  the Hamiltonians $J_{\pm}$ are expressed as
\[
J_+=\sum_{k=1}^N (c_k+d_k), \quad
J_-=\sum_{k=1}^N\frac{c_k+d_k}{d_kd_{k+1}},
\]
and the corresponding Hamiltonian flows read:
\begin{equation}
\dot{c}_k=c_k(d_{k+1}+c_{k+1}-d_k-c_{k-1}), \quad
\dot{d}_k=d_k(c_k-c_{k-1})
\end{equation}
for the $J_+$ Hamiltonian, and
\begin{equation}
\dot{c}_k=c_k\left(\frac{1}{d_k}-\frac{1}{d_{k+1}}\right), \quad
\dot{d}_k=d_k\left(\frac{c_k}{d_kd_{k+1}}-
\frac{c_{k-1}}{d_{k-1}d_k}\right)
\end{equation}
for the $J_-$ Hamiltonian.

The fundamental matrices $L,U$ have in new coordinates the form
\begin{equation}
L=\sum_{k=1}^N d_kE_{kk}+\lambda\sum_{k=1}^N E_{k+1,k},
\end{equation}
\begin{equation}
U=\sum_{k=1}^N E_{kk}-\lambda^{-1}\sum_{k=1}^N c_kE_{k,k+1}.
\end{equation}
For the further reference we give here also the expressions in the
variables $c_k,d_k$ for the matrices involved in the theorem 1:
\begin{eqnarray}
A(c,d,\lambda) & = & \sum_{k=1}^N(d_k+c_{k-1})E_{kk}+\lambda
\sum_{k=1}^N E_{k+1,k},\\
B(c,d,\lambda) & = & \sum_{k=1}^N(d_k+c_k)E_{kk}+\lambda
\sum_{k=1}^N E_{k+1,k},\\
C(c,d,\lambda) & = & -\lambda^{-1}\sum_{k=1}^N
\frac{c_k}{d_{k+1}}E_{k,k+1}, \\
D(c,d,\lambda) & = & -\lambda^{-1}\sum_{k=1}^N
\frac{c_k}{d_k}E_{k,k+1}.
\end{eqnarray}

\setcounter{equation}{0}
\section{Algebraic structure}
Here we recall some of the results of [11,12] on the algebraic interpretation
of the relativistic Toda lattice as a Hamiltonian system on a particular
orbit of a certain Poisson bracket on a matrix group ([11] deals with
 a gauge transformed Lax matrix, which results in a different
Poisson bracket on a group).
The results concerning the difference equations (part c of the Theorem 2
below) are, to my knowledge, new; however the similar results for less
general Poisson brackets can be found in [13,14].

First of all, we define the relevant algebras, groups and decompositions.

1) For the open--end case we set ${\rm\bf g}=gl(N)$. As a linear space,
${\rm\bf g}$ may be represented as a direct sum of two subspaces, which
serve also as subalgebras: ${\rm\bf g}={\rm\bf g}_+\oplus{\rm\bf g}_-$,
where ${\rm\bf g}_+$ (${\rm\bf g}_-$) is a space of all lower triangular
(resp. strictly upper triangular)  $N$ by $N$ matrices. The corresponding
groups are: ${\rm\bf G}=GL(N)$; ${\rm\bf G}_+$ (${\rm\bf G}_-$) is a group
of all nondegenerate lower triangular $N$ by $N$ matrices (resp. upper
triangular $N$ by $N$ matrices with unities on the diagonal).

2) For the periodic case ${\rm\bf g}$ is a certain twisted loop algebra
over $gl(N)$:
\[
{\rm\bf g}=\left\{\tau(\lambda)\in gl(N)[\lambda,\lambda^{-1}]:
\Omega\tau(\lambda)\Omega^{-1}=\tau(\omega\lambda)\right\},
\]
where $\Omega={\rm diag}(1,\omega,\ldots,\omega^{N-1})$,
$\omega=\exp(2\pi i/N)$. Again, as a linear space ${\rm\bf g}=
{\rm\bf g}_+\oplus{\rm\bf g}_-$, where ${\rm\bf g}_+$ (${\rm\bf g}_-$)
is a subspace and subalgebra consisting of $\tau(\lambda)$ containing
only non--negative (resp. only negative) powers of $\lambda$. The
corresponding groups are: ${\rm\bf G}$, the twisted loop group, i.e.
the group of $GL(N)$--valued functions $T(\lambda)$ of the complex
parameter $\lambda$,
regular in ${\Bbb C}P^1\backslash\{0,\infty\}$ and satisfying $\Omega
T(\lambda)
\Omega^{-1}=T(\omega\lambda)$; ${\rm\bf G}_+$ (${\rm\bf G}_-$) is the
subgroup consisting of $T(\lambda)$ regular in the neighbourhood of
$\lambda=0$ (resp. regular in the neighbourhood of $\lambda=\infty$ and
taking the value $I$ in $\lambda=\infty$).

For both the open--end and periodic cases every $\tau\in{\rm\bf g}$ admits
a unique decomposition $\tau=l-u$, where $l\in{\rm\bf g}_+$, $u\in
{\rm\bf g}_-$. We denote $l=\pi_+(\tau)$, $u=\pi_-(\tau)$. Analogously,
for the both cases every $T\in{\rm\bf G}$ from some
neighbourhood of the group unity admits a unique factorization
$T={\cal L}\,{\cal U}^{-1}$, where
${\cal L}\in{\rm\bf G}_+$, ${\cal U}\in{\rm\bf G}_-$. We denote the
factors as ${\cal L}=\Pi_+(T)$, ${\cal U}=\Pi_-(T)$.

Recall also that the derivative $d\varphi(T)$ of the conjugation invariant
function $\varphi: {\rm\bf G}\mapsto{\Bbb C}$ is defined by the relation
\[
{\rm tr}(d\varphi(T)u)=\left.\frac{d}{d\varepsilon}\varphi(Te^
{{\displaystyle\varepsilon u}})\right|_{\varepsilon=0}
=\left.\frac{d}{d\varepsilon}\varphi(
e^{{\displaystyle\varepsilon u}}T)\right|_{\varepsilon=0}, \quad
\forall u \in {\rm\bf g}.
\]

{\bf Theorem 2.}

a) {\it Equip ${\rm\bf G}\times{\rm\bf G}$ with the quadratic Poisson
bracket  (38)--(41) from Ref. {\rm [12]}, and ${\rm\bf G}$ with the quadratic
Poisson bracket (33) from Ref. {\rm [12]}. Then the set
of pairs of matrices $\{(L(c,d,\lambda),U(c,d,\lambda))\}$
forms a Poisson submanifold in  ${\rm\bf G}\times{\rm\bf G}$,
the set of matrices $\{T_{\pm}(c,d,\lambda)\}$ forms a Poisson
submanifold in ${\rm\bf G}$, and the maps
$(L,U)\mapsto T_+=LU^{-1}$ and $(L,U)\mapsto T_-=U^{-1}L$ are Poisson maps
from ${\rm\bf G}\times{\rm\bf G}$ into ${\rm\bf G}$.

{\rm b)} Let $\varphi: {\rm\bf G}\mapsto{\Bbb C}$ be an invariant function
on ${\rm\bf G}$. Then the Hamiltonian flow on ${\rm\bf G}\times{\rm\bf G}$
with the Hamiltonian function
$\varphi(LU^{-1})=\varphi(U^{-1}L)$ has the form
\begin{eqnarray*}
\dot{L} & = &L\pi_{\pm}(d\varphi(T_-))-\pi_{\pm}(d\varphi(T_+))L, \\
\dot{U} & = &U\pi_{\pm}(d\varphi(T_-))-\pi_{\pm}(d\varphi(T_+))U,
\end{eqnarray*}
and the Hamiltonian flow on ${\rm\bf G}$ with the Hamiltonian function
$\varphi(T)$ has the form
\[
\dot{T}=\left[T,\pi_{\pm}(d\varphi(T))\right], \quad T=T_+ \quad {\rm or}
\quad T_- .
\]
These flows admit the following solution in terms of the factorization
problem
\[
e^{{\displaystyle td\varphi(T_{\pm}(0))}}=
{\cal L}_{\pm}(t)\,{\cal U}^{-1}_{\pm}(t),\quad
{\cal L}_{\pm}(t)\in {\rm\bf G}_+,
\quad {\cal U}_{\pm}(t)\in {\rm\bf G}_-
\]
(this problem has solutions at least for sufficiently small t):
\[
L(t)={\cal L}_+^{-1}(t)L(0){\cal L}_-(t)=
{\cal U}_+^{-1}(t)L(0)\,{\cal U}_-(t),
\]
\[
U(t)={\cal L}_+^{-1}(t)U(0){\cal L}_-(t)=
{\cal U}_+^{-1}(t)U(0)\,{\cal U}_-(t),
\]
so that
\[
T_{\pm}(t)={\cal L}_{\pm}^{-1}(t)T_{\pm}(0){\cal L}_{\pm}(t)=
{\cal U}_{\pm}^{-1}(t)T_{\pm}(0)\,{\cal U}_{\pm}(t).
\]

{\rm c)} If $f:{\rm\bf G}\mapsto{\rm\bf G}$ is the derivative of an
invariant function on ${\rm\bf G}$, then the system of difference
equations ($t\in h{\Bbb Z}$)
\[
L(t+h)=\Pi_{\pm}^{-1}\Big(f(T_+(t))\Big)L(t)
\Pi_{\pm}\Big(f(T_-(t))\Big),
\]
\[
U(t+h)=\Pi_{\pm}^{-1}\Big(f(T_+(t))\Big)U(t)
\Pi_{\pm}\Big(f(T_-(t))\Big)
\]
defines a Poisson map ${\rm\bf G}\times{\rm\bf G}\mapsto
{\rm\bf G}\times{\rm\bf G}$, and the difference equation
\[
T(t+h)=\Pi_{\pm}^{-1}\Big(f(T(t))\Big)T(t)
\Pi_{\pm}\Big(f(T(t))\Big),\quad T=T_+ \quad {\rm or}
\quad T_-
\]
defines a Poisson map ${\rm\bf G}\mapsto{\rm\bf G}$. These difference
equations admit following solution in terms of the factorization problem
\[
f^n(T_{\pm}(0))=
{\cal L}_{\pm}(nh)\,{\cal U}^{-1}_{\pm}(nh),
\quad {\cal L}_{\pm}(nh)\in {\rm\bf G}_+,
\quad {\cal U}_{\pm}(nh)\in {\rm\bf G}_-
\]
(this problem has solutions for a given $n$ at least if $f(T_{\pm}(0))$ is
sufficiently close to the group unity $I$):
\[
L(nh)={\cal L}_+^{-1}(nh)L(0){\cal L}_-(nh)=
{\cal U}_+^{-1}(nh)L(0)\,{\cal U}_-(nh),
\]
\[
U(nh)={\cal L}_+^{-1}(nh)U(0){\cal L}_-(nh)=
{\cal U}_+^{-1}(nh)U(0)\,{\cal U}_-(nh),
\]
so that
\[
T_{\pm}(nh)={\cal L}_{\pm}^{-1}(nh)T_{\pm}(0){\cal L}_{\pm}(nh)=
{\cal U}_{\pm}^{-1}(nh)T_{\pm}(0)\,{\cal U}_{\pm}(nh).
\]

{\rm d)} The solutions of the difference equations of the part {\rm c)} are
interpolated by the flows of the part {\rm b)} with the Hamiltonian function
$\varphi(T)$ defined by
\[
d\varphi(T)=h^{-1}\log(f(T)).
\]
}

The part b) of the last theorem explains, in particular, the theorem 1, as
for $J_+(T)={\rm tr}(T)$, $J_-(T)={\rm tr}(T^{-1})$ we have
\[
dJ_+(T)=T, \quad dJ_-(T)=-T^{-1},
\]
and it is not hard to check that
\[
A=\pi_+(T_-), \quad B=\pi_+(T_+), \quad C=\pi_-(-T_-^{-1}), \quad
D=\pi_-(-T_+^{-1}).
\]

\setcounter{equation}{0}
\section{Discrete equations of motion and
\newline symplectic structure}

We propose two systems of difference equations as  discrete--time versions of
the relativistic Toda latti
ce. The equations of motion of the first system read:
\[
\frac{\exp(x_k(t+h)-x_k(t))-1}{\exp(x_k(t)-x_k(t-h))-1}=
\]
\begin{equation}
\frac{\Big(1+g^2\exp(x_{k+1}(t)-x_k(t))\Big)}
{\Big(1+g^2\exp(x_{k+1}(t-h)-x_k(t))\Big)}
\frac{\Big(1+g^2\exp(x_k(t)-x_{k-1}(t+h))\Big)}
{\Big(1+g^2\exp(x_k(t)-x_{k-1}(t))\Big)},
\quad 1\le k \le N.
\end{equation}
The equations of motion of the second one read:
\[
\frac{\exp(-x_k(t+h)+x_k(t))-1}{\exp(-x_k(t)+x_k(t-h))-1}=
\]
\begin{equation}
\frac{\Big(1+g^2\exp(x_{k+1}(t+h)-x_k(t))\Big)}
{\Big(1+g^2\exp(x_{k+1}(t)-x_k(t))\Big)}
\frac{\Big(1+g^2\exp(x_k(t)-x_{k-1}(t))\Big)}
{\Big(1+g^2\exp(x_k(t)-x_{k-1}(t-h))\Big)},
\quad 1\le k \le N,
\end{equation}
The both systems are considered under one of the two types of boundary
conditions, either open--end:
\[
x_0(t)\equiv\infty,\quad x_{N+1}(t)\equiv-\infty \quad {\rm for \quad all}\quad
t\in
h{\Bbb Z},
\]
or periodic:
\[
x_0(t)\equiv x_N(t),\quad x_{N+1}(t)\equiv x_1(t) \quad {\rm for \quad
all}\quad t\in
h{\Bbb Z}.
\]
The functions $x_k(t)$ in (4.1), (4.2) are supposed to be defined for $t\in h
{\Bbb Z}$, $h>0$ (note that the equations themselves do not depend on $h$
explicitly). However, it is convenient to consider (4.1), (4.2)
as  finite--difference approximations
to (2.1), and in this context $x_k(t)$ are to be considered as the smooth
functions of $t\in{\Bbb R}$. Then the left--hand sides are expanded in
powers of $h$ as $1+h\displaystyle\frac{\ddot{x}_k}{\dot{x}_k}+O(h^2)$, and the
 right--hand sides as
\[
1+h\left(\dot{x}_{k+1}\frac{g^2\exp(x_{k+1}-x_k)}{1+g^2\exp(x_{k+1}-x_k)}-
\dot{x}_{k-1}\frac{g^2\exp(x_k-x_{k-1})}{1+g^2\exp(x_k-x_{k-1})}\right)+O(h^2),
\]
so that we recover in the continuous limit the equation (2.1).

Of course, the both systems are closely related, namely by means of the time
reversion operation. (Note that the underlying continuous--time system (2.1)
is invariant with respect to this operation).

We would like to remark that the equations (4.1), (4.2) admit a simple
non--relativistic limit: set $x_k(t)=q_k(t)+ct$ in (4.1)
(resp. $x_k(t)=q_k(t)-ct$ in (4.2)) with $c>0$ playing the role of the
speed of light;
then in the limit $c\to\infty$ the both equations tend to one and the
same system:
\[
\exp(q_k(t+h)-2q_k(t)+q_k(t-h))=
\frac{1+g^2\exp(q_{k+1}(t)-q_k(t))}{1+g^2\exp(q_k(t)-q_{k-1}(t))},
\quad 1\le k \le N,
\]
i.e. to the equations of motion of the discrete--time Toda lattice from [3].

\renewcommand{\arraystretch}{0.7}

In the following we will adopt the notations from [7,8]: if $x_k=x_k(t)$,
then $\wx_k=x_k(t+h)$, $\uxk=x_k(t-h)$, so that (4.1), (4.2) take the form

\begin{equation}
\frac{\exp(\wx_k-x_k)-1}{\exp(x_k-\uxk)-1}=
\frac{(1+g^2\exp(x_{k+1}-x_k))}{(1+g^2\exp(\uxp-x_k))}\:
\frac{(1+g^2\exp(x_k-\wx_{k-1}))}{(1+g^2\exp(x_k-x_{k-1}))},
\quad 1\le k \le N;
\end{equation}

\begin{equation}
\frac{\exp(-\wx_k+x_k)-1}{\exp(-x_k+\uxk)-1}=
\frac{(1+g^2\exp(\wx_{k+1}-x_k))}{(1+g^2\exp(x_{k+1}-x_k))}\:
\frac{(1+g^2\exp(x_k-x_{k-1}))}{(1+g^2\exp(x_k-\uxm))},
\quad 1\le k \le N,
\end{equation}
respectively. Obviously, (4.3), (4.4) are systems of nonlinear algebraic
equations for $\wx_k$, $1\le k \le N$ of the form
\begin{equation}
F_k(\wx,x,\ux)=0,\quad 1\le k \le N.
\end{equation}
 It will be convenient to discuss the solvability of these systems
simultaneously with the symplectic structure.

The general recipe to derive the invariant symplectic structure for a
discrete evolutionary equation of the type (4.5)
was given in [1,15]: represent the equations
of motion in the Lagrangian form
\begin{equation}
\partial\left(\Lambda(\wx,x)+\Lambda(x,\ux)\right)/\partial x_k=0,
\end{equation}
then the momenta $p_k$ canonically conjugate to $x_k$ are defined as
\begin{equation}
p_k=\partial\Lambda(x,\ux)/\partial x_k.
\end{equation}
The map $(x,p)\mapsto (\wx,\widetilde{p})$ preserves the standard
symplectic 2--form $\sum_{k=1}^Ndx_k\wedge dp_k$.
Note that for the $\wx$ we have the equation
\begin{equation}
p_k=-\partial\Lambda(\wx,x)/\partial x_k,
\end{equation}
and then  $\widetilde{p}$ may be computed as
\begin{equation}
\widetilde{p}_k=\partial\Lambda(\wx,x)/\partial \widetilde{x}_k.
\end{equation}

{\bf Remark.} Note that if the system (4.5) may be represented in the
Lagrangian
form (4.6) with the Lagrangian function $\Lambda(x,\ux)$,
then the system $F_k(\ux,x,\wx)=0$, $1\le k \le N$ may be represented in
the Lagrangian form with the Lagrangian function $-\Lambda(\ux,x)$.
Now it is easy to see from (4.7)--(4.9) that the symplectic maps
$(x,p)\mapsto (\wx,\widetilde{p})$ corresponding to these two systems are
mutually inverse. This is just the case for (4.1), (4.2), up to minor
modifications due to the "artificial" parameter $h$ introduced
in the definitions of momenta below. We prefer, however, to consider all the
maps separately, and will return to the above--mentioned circumstance
at the end of the section 6.

The Lagrangian functions for the equations (4.3), (4.4) are expressed with the
help of two functions $\phi_1(\xi), \phi_2(\xi)$ which are defined by
\[
\phi_1^{\prime}(\xi)=\log\left|\frac{\exp(\xi)-1}{h}\right|,\quad
\phi_2^{\prime}(\xi)=\log(1+g^2\exp(\xi)).
\]
It turns out that, just as in the continuous--time case, the Lagrangian
functions
and hence the momenta $p_k$ may be choosen in two different ways. They lead to
two pairs of different symplectic maps (one pair for each of (4.1), (4.2))
 belonging, remarkably, to one and the same
integrable hierarchy. Still more remarkable, however, is that this
hierarchy is the same  as in the continuous--time case (see sections 5, 6)!

\subsection{System (4.3), the first choice of momenta}
It is easy to see that (4.3) is equivalent to (4.6) with
\[
\Lambda(x,\ux)=\sum_{k=1}^N\phi_1(x_k-\uxk)+\sum_{k=1}^N\phi_2(\uxk-x_{k-1})-
\sum_{k=1}^N\phi_2(\uxk-\uxm).
\]
Then the definition (4.7) of momenta $p_k$ takes the form
\begin{equation}
\exp(p_k)=\frac{(\exp(x_k-\uxk)-1)}{h(1+g^2\exp(\uxp-x_k))},
\end{equation}
and the equation (4.8) for $\wx$ takes the form
\begin{equation}
\exp(p_k)=\frac{(\exp(\wx_k-x_k)-1)}{h(1+g^2\exp(x_{k+1}-x_k))}\:
\frac{(1+g^2\exp(x_k-x_{k-1}))}{(1+g^2\exp(x_k-\wx_{k-1}))},
\end{equation}
(formulas (4.10), (4.11) are to be compared with (2.2)).
The equation (4.9) for $\widetilde{p}$ together with (4.11) implies :
\begin{equation}
\exp(\widetilde{p}_k)=\exp(p_k)\:
\frac{(1+g^2\exp(x_{k+1}-x_k))}{(1+g^2\exp(x_{k+1}-\wx_k))}\:
\frac{(1+g^2\exp(x_k-\wx_{k-1}))}{(1+g^2\exp(x_k-x_{k-1}))}.
\end{equation}

To solve (4.3) for $\wx$ is now equivalent to  solving
 (4.11) for $\wx$. It may be directly verified that this last equation
 may be rewritten as:
\[
h\exp(p_k)\;\frac{1+g^2\exp(x_{k+1}-\wx_k)}{1-\exp(x_k-\wx_k)}=
\]
\[
=1+h\exp(p_k)+g^2\exp(x_k-x_{k-1})\frac{1-\exp(x_{k-1}-\wx_{k-1})}
{1+g^2\exp(x_k-\wx_{k-1})}.
\]
In terms of the coordinates $c_k, d_k$ this means: if we denote
\begin{equation}
{\goth a}_k=h\exp(p_k)\;\frac{1+g^2\exp(x_{k+1}-\wx_k)}{1-\exp(x_k-\wx_k)},
\end{equation}
then
\begin{equation}
{\goth a}_k=1+hd_k+\frac{hc_{k-1}}{{\goth a}_{k-1}}, \quad 1\le k \le N.
\end{equation}
Suppose for a moment that these recurrent relations define ${\goth a}_k$ as
certain functions of $c_k,d_k$. Then, according to (4.13), this means that
the equations for $\wx$ are solved. Indeed, it follows from (4.13) that
\[
\exp(\wx_k-x_k)=\frac{{\goth a}_k+hc_k}{{\goth a}_k-hd_k}
\]
To express now the resulting map in terms of the variables $c_k, d_k$ alone,
we derive from (4.13)
\[
\frac{1+g^2\exp(x_k-x_{k-1})}{1+g^2\exp(x_k-\wx_{k-1})}=
{\goth a}_k-hd_k,
\]
which together with the previous expression and (4.12) implies:
\begin{equation}
\widetilde{c}_k=c_k\frac{{\goth a}_{k+1}+hc_{k+1}}{{\goth a}_k+hc_k},\quad
\widetilde{d}_k=d_k\frac{{\goth a}_{k+1}-hd_{k+1}}{{\goth a}_k-hd_k}.
\end{equation}
Returning to the relations (4.14), we note that in the open--end case
$c_0=0$, hence we obtain from (4.11) the following
finite continued fractions expressions for ${\goth a}_k$'s:
\[
{\goth a}_1=1+hd_1;
\]
\[
{\goth a}_2=1+hd_2+\frac{hc_1}{1+hd_1};
\]
\[
\cdots
\]
\[
{\goth a}_N=1+hd_N+\frac{hc_{N-1}}{1+hd_{N-1}+
\displaystyle\frac{hc_{N-2}}{1+hd_{N-2
}+
\parbox[t]{0.7cm}{$\begin{array}{c}\\  \ddots\end{array}$}
\parbox[t]{2cm}{$\begin{array}{c}
 \\  \\+\displaystyle\frac{hc_1}{1+hd_1}\end{array}$}}}.
\]
Obviously, we have:
\begin{equation}
{\goth a}_k=1+h(d_k+c_{k-1})+O(h^2).
\end{equation}
In the periodic case the recurrent relations (4.14) uniquely define the
${\goth a}_k$'s as the $N$-periodic infinite continued fractions. It can be
proved that these continued fractions converge and their values satisfy
 (4.16).

Because of (4.16) it is obvious that the map (4.15) is a difference
approximation to the flow (2.8).

\subsection{System (4.3), the second choice of momenta}
The equation (4.3) can be treated also as a Lagrangian equation (4.6) with
\[
\Lambda(x,\ux)=-\sum_{k=1}^N\phi_1(x_k-\uxk)-\sum_{k=1}^N\phi_2(\uxk-x_{k-1})+
\sum_{k=1}^N\phi_2(x_k-x_{k-1}).
\]
Then the definition (4.7) takes the form:
\begin{equation}
\exp(p_k)=\frac{h(1+g^2\exp(x_k-x_{k-1}))}{(1-\exp(x_k-\uxk))}\:
\frac{(1+g^2\exp(\uxp-x_k))}{(1+g^2\exp(x_{k+1}-x_k))}.
\end{equation}
and the equation (4.8) for $\wx$ takes the form:
\begin{equation}
\exp(p_k)=\frac{h(1+g^2\exp(x_k-\wx_{k-1}))}{1-\exp(\wx_k-x_k)},
\end{equation}
(Equations (4.17), (4.18) are to be compared with (2.4)).
Equation (4.9) together with (4.18) implies::
\begin{equation}
\exp(\widetilde{p}_k)=\exp(p_k)\:
\frac{(1+g^2\exp(x_{k+1}-\wx_k))}{(1+g^2\exp(\wx_{k+1}-\wx_k))}\:
\frac{(1+g^2\exp(\wx_k-\wx_{k-1}))}{(1+g^2\exp(x_k-\wx_{k-1}))}.
\end{equation}
As before, to solve (4.3) for $\wx$ is  equivalent to solving
 (4.18) for $\wx$. The formula (4.18) may be rewritten as:
\[
\exp(\wx_k-x_k+p_k)=\exp(p_k)-h-hg^2\exp(x_k-\wx_{k-1}).
\]
In terms of the coordinates $c_k, d_k$ this means: if we denote
\begin{equation}
{\goth d}_k=g^2\exp(x_{k+1}-\wx_k),
\end{equation}
then
\begin{equation}
\frac{c_k}{{\goth d}_k}=d_k-h-h{\goth d}_{k-1}, \quad 1\le k \le N.
\end{equation}

Supposing that thes relations define ${\goth d}_k$'s as certain functions on
$c_k,d_k$, we see that (4.20) allows to solve the equation for $\wx$.
In order to express the resulting map in terms of the variables $c_k, d_k$
alone, we derive from (4.18), (4.20) the relations
\[
\exp(x_k-\wx_k)=\frac{d_k{\goth d}_k}{c_k},\quad
\frac{1+g^2\exp(\wx_{k+1}-\wx_k)}{1+g^2\exp(x_{k+1}-\wx_k)}=
1-\displaystyle\frac{h{\goth d}_k}{d_{k+1}}.
\]
We use them and (4.21) to derive from (4.19):
\begin{equation}
\widetilde{c}_k=
c_{k+1}\frac{c_k+h{\goth d}_k}{c_{k+1}+h{\goth d}_{k+1}}, \quad
\widetilde{d}_k=
d_{k+1}\frac{d_k-h{\goth d}_{k-1}}{d_{k+1}-h{\goth d}_k}.
\end{equation}

Returning to (4.21), we immediately obtain in the open--end case ${\goth
d}_0=0$,
and the finite continued fraction expressions for ${\goth d}_k$'s follow:
\[
{\goth d}_1=\frac{c_1}{d_1-h},
\]
\[
{\goth d}_2=\frac{c_2}{d_2-h-\displaystyle\frac{hc_1}{d_1-h}},
\]
\[
\cdots
\]
\[
{\goth d}_{N-1}=\frac{c_{N-1}}{d_{N-1}-h-\displaystyle
\frac{hc_{N-2}}{d_{N-2}-h-
\parbox[t]{0.7cm}{$\begin{array}{c}\\  \ddots\end{array}$}
\parbox[t]{2cm}{$\begin{array}{c}
 \\  \\-\displaystyle\frac{hc_1}{d_1-h}\end{array}$}}}.
\]
Obviously, we have
\begin{equation}
{\goth d}_k=\frac{c_k}{d_k}+O(h),\quad 1\le k \le N.
\end{equation}
In the periodic case the recurrent relations (4.21) uniquely define
${\goth d}_k$'s as the $N$--periodic
infinite continued fractions, that again converge and whose values satisfy
the relation (4.22).
In view of (4.22) it is obvious that the map (4.21) is a difference
approximation to the flow (2.9).

\subsection{System (4.4), the first choice of momenta}
Turning now to the system (4.2), we see that it is equivalent to (4.6) with
\[
\Lambda(x,\ux)=-\sum_{k=1}^N\phi_1(-x_k+\uxk)-\sum_{k=1}^N\phi_2(x_k-\uxm)+
\sum_{k=1}^N\phi_2(x_k-x_{k-1}).
\]
Then the definition (4.7) of momenta $p_k$ takes the form
\begin{equation}
\exp(p_k)=\frac{(1-\exp(-x_k+\uxk))}{h(1+g^2\exp(x_{k+1}-x_k))}\:
\frac{(1+g^2\exp(x_k-x_{k-1}))}{(1+g^2\exp(x_k-\uxm))},
\end{equation}
and the equation (4.8) for $\wx$ takes the form
\begin{equation}
\exp(p_k)=\frac{1-\exp(-\wx_k+x_k)}{h(1+g^2\exp(\wx_{k+1}-x_k))},
\end{equation}
(formulas (4.24), (4.25) are to be compared with (2.2)).
The equation (4.9) for $\widetilde{p}$ together with (4.25) implies :
\begin{equation}
\exp(\widetilde{p}_k)=\exp(p_k)\:
\frac{(1+g^2\exp(\wx_k-\wx_{k-1}))}{(1+g^2\exp(\wx_k-x_{k-1}))}\:
\frac{(1+g^2\exp(\wx_{k+1}-x_k))}{(1+g^2\exp(\wx_{k+1}-\wx_k))}.
\end{equation}

To solve (4.4) for $\wx$ is now equivalent to  solving
(4.25) for $\wx$. This last equation may be rewritten as:
\[
\exp(-\wx_k+x_k)=1-h\exp(p_k)-
\frac{hg^2\exp(x_{k+1}-x_k+p_k)}{\exp(-\wx_{k+1}+x_{k+1})}.
\]
In terms of the coordinates $c_k, d_k$ this means: if we denote
\begin{equation}
\beta_k=\exp(-\wx_k+x_k),
\end{equation}
then
\begin{equation}
\beta_k=1-hd_k-\frac{hc_k}{\beta_{k+1}}, \quad 1\le k \le N.
\end{equation}

Relations (4.28) imply that $\beta_k$ as certain functions (continued
fractions,
see below) on $c_k,d_k$.
According to (4.27), this means that the equations for $\wx$ are solved.
To express now the resulting map in terms of the variables $c_k, d_k$ alone,
we derive from (4.25),(4.27), and (4.28) the equality
\[
\frac{1+g^2\exp(\wx_{k+1}-\wx_k)}{1+g^2\exp(\wx_{k+1}-x_k)}=
1-\frac{hc_k}{\beta_{k+1}}=\beta_k+hd_k,
\]
so that (4.26) implies
\begin{equation}
\widetilde{c}_k=c_k\frac{\beta_k-hc_{k-1}}{\beta_{k+1}-hc_k},\quad
\widetilde{d}_k=d_k\frac{\beta_{k-1}+hd_{k-1}}{\beta_k+hd_k}.
\end{equation}

As for the continued fractions expressions for $\beta_k$'s, we have in
the open--end case: $c_N=0$, hence
\[
\beta_N=1-hd_N;
\]
\[
\beta_{N-1}=1-hd_{N-1}-\frac{hc_{N-1}}{1-hd_N};
\]
\[
\cdots
\]
\[
\beta_1=1-hd_1-\frac{hc_1}{1-hd_2-\displaystyle\frac{hc_2}{1-hd_3-
\parbox[t]{0.7cm}{$\begin{array}{c}\\  \ddots\end{array}$}
\parbox[t]{2cm}{$\begin{array}{c}
 \\  \\-\displaystyle\frac{hc_{N-1}}{1-hd_N}\end{array}$}}}.
\]
Obviously,
\begin{equation}
\beta_k=1-h(c_k+d_k)+O(h^2).
\end{equation}
In the periodic case the recurrent relations (4.28) uniquely define the
$\beta_k$'s as the $N$-periodic infinite continued fractions. It can be
proved that these continued fractions converge and their values satisfy
 (4.30).
Because of (4.30) it is obvious that the map (4.29) is a difference
approximation to the flow (2.8).

\subsection{System (4.4), the second choice of momenta}
The equation (4.4) can be treated also as a Lagrangian equation (4.6) with
\[
\Lambda(x,\ux)=\sum_{k=1}^N\phi_1(-x_k+\uxk)+\sum_{k=1}^N\phi_2(x_k-\uxm)-
\sum_{k=1}^N\phi_2(\uxk-\uxm).
\]
Then the definition (4.7) takes the form:
\begin{equation}
\exp(p_k)=\frac{h(1+g^2\exp(x_k-\uxm))}{\exp(-x_k+\uxk)-1},
\end{equation}
and the equation (4.8) for $\wx$ takes the form:
\begin{equation}
\exp(p_k)=\frac{h(1+g^2\exp(x_k-x_{k-1}))}{(\exp(-\wx_k+x_k)-1)}\:
\frac{(1+g^2\exp(\wx_{k+1}-x_k))}{(1+g^2\exp(x_{k+1}-x_k))}.
\end{equation}
(Equations (4.32), (4.33) are to be compared with (2.4)).
Equation (4.9) together with (4.32) implies::
\begin{equation}
\exp(\widetilde{p}_k)=\exp(p_k)\:
\frac{(1+g^2\exp(\wx_k-x_{k-1}))}{(1+g^2\exp(x_k-x_{k-1}))}\:
\frac{(1+g^2\exp(x_{k+1}-x_k))}{(1+g^2\exp(\wx_{k+1}-x_k))}.
\end{equation}
As before, to solve (4.4) for $\wx$ is  equivalent to solving
 (4.32) for $\wx$. The formula (4.32) may be rewritten as:
\[
\frac{h(1+g^2\exp(\wx_k-x_{k-1}))}{1-\exp(\wx_k-x_k)}=
\]
\[
=h+\exp(p_k)+g^2\exp(x_{k+1}-x_k+p_k)
\frac{1-\exp(\wx_{k+1}-x_{k+1})}{1+g^2\exp(\wx_{k+1}-x_k)}.
\]
In terms of the coordinates $c_k, d_k$ this means: if we denote
\begin{equation}
h\gamma_k=c_k\frac{1-\exp(\wx_{k+1}-x_{k+1})}{1+g^2\exp(\wx_{k+1}-x_k)},
\end{equation}
then
\begin{equation}
\frac{c_{k-1}}{\gamma_{k-1}}=h+d_k+h\gamma_k, \quad 1\le k \le N.
\end{equation}
This defines $\gamma_k$'s as certain functions on $c_k,d_k$,
which, according to (4.34), allows to solve the equation for $\wx$.
It is not hard to derive from (4.34) the relations
\[
\exp(\wx_{k+1}-x_{k+1})=\frac{1-\displaystyle\frac{h\gamma_k}{c_k}}
{1+\displaystyle\frac{h\gamma_k}{d_k}},\quad
\frac{1+g^2\exp(x_{k+1}-x_k)}{1+g^2\exp(\wx_{k+1}-x_k)}=
1+\displaystyle\frac{h\gamma_k}{d_k}.
\]
We use them and (4.33) to
express the resulting map in terms of the variables
 $c_k, d_k$ alone:
\begin{equation}
\widetilde{c}_k=
c_{k-1}\frac{c_k-h\gamma_k}{c_{k-1}-h\gamma_{k-1}},\quad
\widetilde{d}_k=
d_{k-1}\frac{d_k+h\gamma_k}{d_{k-1}+h\gamma_{k-1}}.
\end{equation}
The expressions for $\gamma_k$'s in the open--end case follow from (4.35)
and $\gamma_N=0$:
\[
\gamma_{N-1}=\frac{c_{N-1}}{d_N+h},
\]
\[
\gamma_{N-2}=\frac{c_{N-2}}{d_{N-1}+h+\displaystyle
\frac{hc_{N-1}}{d_N+h}},
\]
\[
\cdots
\]
\[
\gamma_1=\frac{c_1}{d_2+h+\displaystyle
\frac{hc_2}{d_3+h+
\parbox[t]{0.7cm}{$\begin{array}{c}\\  \ddots\end{array}$}
\parbox[t]{2cm}{$\begin{array}{c}
 \\  \\+\displaystyle\frac{hc_{N-1}}{d_N+h}\end{array}$}}}.
\]
Obviously, we have
\begin{equation}
\gamma_k=\frac{c_k}{d_{k+1}}+O(h),\quad 1\le k \le N.
\end{equation}
In the periodic case the recurrent relations (4.35) uniquely define
$\gamma_k$'s as the $N$--periodic
infinite continued fractions, that again converge and whose values satisfy
the relation (4.37).
In view of (4.37) it is obvious that the map (4.36) is a difference
approximation to the flow (2.9).

\setcounter{equation}{0}
\section{Lax representations}
We show in this section that our discrete--time systems admit the discrete
analogs of Lax representations with the same matrices $L,U,T_{\pm}$
(see (2.10), (2.11), (2.6))
 as the continuous--time relativistic Toda lattice. In the
following theorems we adopt the conventions described before the formula
(2.6). The dependence of all the matrices below on the discrete time
$t\in h{\Bbb Z}$ is supposed to
appear through the dependence of $c_k$, $d_k$ on $t$.

{\bf Theorem 3.} {\it
The symplectic map defined by (4.14), (4.15) admits a representation in the
form of matrix equations
\begin{equation}
{\rm\bf A}(t)L(t+h)=L(t){\rm\bf B}(t),\quad {\rm\bf A}(t)U(t+h)=U(t){\rm\bf
B}(t),
\end{equation}
so that
\begin{equation}
T_+(t+h)={\rm\bf A}^{-1}(t)T_+(t){\rm\bf A}(t), \quad
T_-(t+h)={\rm\bf B}^{-1}(t)T_-(t){\rm\bf B}(t)
\end{equation}
with the matrices
\begin{equation}
{\rm\bf A}(c,d,\lambda)=\sum_{k=1}^N{\goth
a}_kE_{kk}+h\lambda\sum_{k=1}^NE_{k+1,k},
\end{equation}
\begin{equation}
{\rm\bf B}(c,d,\lambda)=\sum_{k=1}^N{\goth
b}_kE_{kk}+h\lambda\sum_{k=1}^NE_{k+1,k},
\end{equation}
where ${\goth a}_k$'s are defined by (4.14), and ${\goth b}_k$'s by (5.7) or
(5.8) below.}

{\bf Proof.} It is staightforward to check that the matrix equations (5.1)
are equivalent to the following ones:
\begin{equation}
{\goth a}_k\widetilde{d}_k=d_k{\goth b}_k,\quad
h\widetilde{d}_k+{\goth a}_{k+1}=hd_{k+1}+{\goth b}_k,
\end{equation}
\begin{equation}
{\goth a}_k\widetilde{c}_k=c_k{\goth b}_{k+1},\quad
h\widetilde{c}_k-{\goth a}_{k+1}=hc_{k+1}-{\goth b}_{k+1}.
\end{equation}

Now it would be not hard to check directly that the equations (5.5), (5.6)
are satisfied with the following identifications:
 (4.11), (4.12) for $d_k$, $\widetilde{d}_k$; $c_k=g^2
\exp(x_{k+1}-x_k)d_k$ and similarly for $\widetilde{c}_k$; ,
\[
{\goth a}_k=\exp(\wx_k-x_k)
\frac{(1+g^2\exp(x_{k+1}-\wx_k))}{(1+g^2\exp(x_{k+1}-x_k))}\:
\frac{(1+g^2\exp(x_k-x_{k-1}))}{(1+g^2\exp(x_k-\wx_{k-1}))},
\]
as in (4.13), and ${\goth b}_k=\exp(\wx_k-x_k)$

We prefer, however, to work directly with (5.5), (5.6), not using the
expressions
in terms of the $x_k$ variables. Namely, (5.5) is equivalent to
\begin{equation}
\widetilde{d}_k=d_k\frac{{\goth a}_{k+1}-hd_{k+1}}{{\goth a}_k-hd_k},\quad
{\goth b}_k={\goth a_k}\frac{{\goth a}_{k+1}-hd_{k+1}}{{\goth a}_k-hd_k},
\end{equation}
and (5.6) is equivalent to
\begin{equation}
\widetilde{c}_k=c_k\frac{{\goth a}_{k+1}+hc_{k+1}}{{\goth a}_k+hc_k}, \quad
{\goth b}_{k+1}={\goth a}_k\frac{{\goth a}_{k+1}+hc_{k+1}}{{\goth a}_k+hc_k}.
\end{equation}
The first equations in (5.7), (5.8) coincide with (4.15), and the
compatibility of the second ones is equivalent to
\[
\frac{{\goth a}_k({\goth a}_{k+1}-hd_{k+1})}{{\goth a}_k+hc_k}=
\frac{{\goth a}_{k-1}({\goth a}_k-hd_k)}{{\goth a}_{k-1}+hc_{k-1}},
\]
which is a direct consequence of (4.14). This completes the proof.

We would like to mention here that the matrices (5.3), (5.4) when compared with
(2.12), (2.13) satisfy
\[
{\rm\bf A}(c,d,\lambda)=I+hA(c,d,\lambda)+O(h^2),\quad
{\rm\bf B}(c,d,\lambda)=I+hB(c,d,\lambda)+O(h^2),
\]
as it follows from (4.16).

{\bf Theorem 4.} {\it
The symplectic map defined by  (4.21), (4.22) admits a representation in the
form of matrix equations
\begin{equation}
L(t+h){\rm\bf D}(t)={\rm\bf C}(t)L(t),\quad U(t+h){\rm\bf D}(t)={\rm\bf
C}(t)U(t),
\end{equation}
so that
\begin{equation}
T_+(t+h)={\rm\bf C}(t)T_+(t){\rm\bf C}^{-1}(t), \quad
T_-(t+h)={\rm\bf D}(t)T_-(t){\rm\bf D}^{-1}(t)
\end{equation}
with the matrices
\begin{equation}
{\rm\bf C}(c,d,\lambda)=\sum_{k=1}^NE_{kk}+
h\lambda^{-1}\sum_{k=1}^N{\goth c}_kE_{k,k+1},
\end{equation}
\begin{equation}
{\rm\bf D}(c,d,\lambda)=\sum_{k=1}^NE_{kk}+
h\lambda^{-1}\sum_{k=1}^N{\goth d}_kE_{k,k+1},
\end{equation}
where ${\goth d}_k$'s are defined by (4.21), and ${\goth c}_k$'s by (5.15) or
(5.16) below.}

{\bf Proof.} It is staightforward to check that the matrix equations (5.9) are
equivalent to the following ones:
\begin{equation}
\widetilde{d}_k{\goth d}_k={\goth c}_kd_{k+1},\quad
\widetilde{d}_k+h{\goth d}_{k-1}=d_k+h{\goth c}_k,
\end{equation}
\begin{equation}
\widetilde{c}_k{\goth d}_{k+1}={\goth c}_kc_{k+1},\quad
\widetilde{c}_k-h{\goth d}_k=c_k-h{\goth c}_k.
\end{equation}

Now it could be checked by means of direct calculation that the equations
(5.13), (5.14) are satisfied
with the following identifications: (4.18), (4.19) for $d_k$,
$\widetilde{d}_k$; $c_k=g^2
\exp(x_{k+1}-x_k)d_k$ and similarly for $\widetilde{c}_k$;
${\goth d}_k=g^2\exp(x_{k+1}-\wx_k)$ as in (4.2), and
\[
{\goth c}_k=g^2\exp(x_{k+1}-\wx_k)
\frac{(1-\exp(\wx_{k+1}-x_{k+1}))}{(1-\exp(\wx_k+-x_k))}\:
\frac{(1+g^2\exp(\wx_k-\wx_{k-1}))}{(1+g^2\exp(\wx_{k+1}-\wx_k))},
\].

Working, however, again directly with (5.13), (5.14), without turning to the
representation through the $x_k$ variables, we note that (5.13) is equivalent
to
\begin{equation}
\widetilde{d}_k=d_{k+1}\frac{d_k-h{\goth d}_{k-1}}{d_{k+1}-h{\goth d}_k},\quad
{\goth c}_k={\goth d}_k\frac{d_k-h{\goth d}_{k-1}}{d_{k+1}-h{\goth d}_k},
\end{equation}
and (5.14) is equivalent to
\begin{equation}
\widetilde{c}_k=c_{k+1}\frac{c_k+h{\goth d}_k}{c_{k+1}+h{\goth d}_{k+1}},\quad
{\goth c}_k={\goth d}_{k+1}\frac{c_k+h{\goth d}_k}{c_{k+1}+h{\goth d}_{k+1}}.
\end{equation}
The first equations in (5.15), (5.16) coincide with (4.22), and the
compatibility of the second ones is equivalent to
\[
\frac{{\goth d}_{k+1}(d_{k+1}-h{\goth d}_k)}{c_{k+1}+h{\goth d}_{k+1}}=
\frac{{\goth d}_k(d_k-h{\goth d}_{k-1})}{c_k+h{\goth d}_k},
\]
which is a direct consequence of (4.21). The proof is complete.

Note that the matrices (5.11), (5.12) when compared with
(2.14), (2.15) satisfy
\[
{\rm\bf C}(c,d,\lambda)=I-hC(c,d,\lambda)+O(h^2),\quad
{\rm\bf D}(c,d,\lambda)=I-hD(c,d,\lambda)+O(h^2),
\]
as it follows from (4.23).

{\bf Theorem 5.} {\it
The symplectic map defined by (4.28), (4.29) admits a representation in the
form of matrix equations
\begin{equation}
L(t+h){\cal B}(t)={\cal A}(t)L(t),\quad U(t+h){\cal B}(t)={\cal A}(t)U(t),
\end{equation}
so that
\begin{equation}
T_+(t+h)={\cal A}(t)T_+(t){\cal A}^{-1}(t), \quad
T_-(t+h)={\cal B}(t)T_-(t){\cal B}^{-1}(t)
\end{equation}
with the matrices
\begin{equation}
{\cal A}(c,d,\lambda)=\sum_{k=1}^N\alpha_kE_{kk}-h\lambda\sum_{k=1}^NE_{k+1,k},
\end{equation}
\begin{equation}
{\cal B}(c,d,\lambda)=\sum_{k=1}^N\beta_kE_{kk}-h\lambda\sum_{k=1}^NE_{k+1,k},
\end{equation}
where $\beta_k$'s are defined by (4.28), and $\alpha_k$'s by (5.23) or (5.24)
below.}

{\bf Proof.} It is staightforward to check that the matrix equations (5.17)
are equivalent to the following ones:
\begin{equation}
\widetilde{d}_k\beta_k=\alpha_kd_k,\quad h\widetilde{d}_k-\beta_{k-1}=
hd_{k-1}-\alpha_k,
\end{equation}
\begin{equation}
\widetilde{c}_k\beta_{k+1}=\alpha_kc_k,\quad h\widetilde{c}_k+\beta_k=
hc_{k-1}+\alpha_k.
\end{equation}

Now it would be not hard to check directly that the equations (5.5), (5.6)
are satisfied with the following identifications:
 (4.25), (4.26) for $d_k$, $\widetilde{d}_k$; $c_k=g
^2
\exp(x_{k+1}-x_k)d_k$ and similarly for $\widetilde{c}_k$; $\beta_k=
\exp(-\wx_k+x_k)$ as in (4.27), and
\[
\alpha_k=\exp(-\wx_k+x_k)
\frac{(1+g^2\exp(\wx_{k+1}-x_k))}{(1+g^2\exp(\wx_{k+1}-\wx_k))}\:
\frac{(1+g^2\exp(\wx_k-\wx_{k-1}))}{(1+g^2\exp(\wx_k-x_{k-1}))}.
\]

We prefer, however, to work directly with (5.21), (5.22), not using the
expressions
in terms of the $x_k$ variables. Namely, (5.21) is equivalent to
\begin{equation}
\widetilde{d}_k=d_k\frac{\beta_{k-1}+hd_{k-1}}{\beta_k+hd_k},\quad
\alpha_k=\beta_k\frac{\beta_{k-1}+hd_{k-1}}{\beta_k+hd_k},
\end{equation}
and (5.22) is equivalent to
\begin{equation}
\widetilde{c}_k=c_k\frac{\beta_k-hc_{k-1}}{\beta_{k+1}-hc_k}, \quad
\alpha_k=\beta_{k+1}\frac{\beta_k-hc_{k-1}}{\beta_{k+1}-hc_k}.
\end{equation}
The first equations in (5.23), (5.24) coincide with (4.29),  and the
compatibility of the second ones is equivalent to
\[
\frac{\beta_{k+1}(\beta_k+hd_k)}{\beta_{k+1}-hc_k}=
\frac{\beta_k(\beta_{k-1}+hd_{k-1})}{\beta_k-hc_{k-1}},
\]
which is a direct consequence of (4.28). This completes the proof.

We would like to mention here that the matrices (5.19), (5.20) when compared
with
(2.12), (2.13) satisfy
\[
{\cal A}(c,d,\lambda)=I-hA(c,d,\lambda)+O(h^2),\quad
{\cal B}(c,d,\lambda)=I-hB(c,d,\lambda)+O(h^2),
\]
as it follows from (4.30).

{\bf Theorem 6.} {\it
The symplectic map defined by (4.35), (4.36) admits a representation in the
form of matrix equations
\begin{equation}
{\cal C}(t)L(t+h)=L(t){\cal D}(t),\quad {\cal C}(t)U(t+h)=U(t){\cal D}(t),
\end{equation}
so that
\begin{equation}
T_+(t+h)={\cal C}^{-1}(t)T_+(t)\,{\cal C}(t), \quad
T_-(t+h)={\cal D}^{-1}(t)T_-(t){\cal D}(t)
\end{equation}
with the matrices
\begin{equation}
{\cal C}(c,d,\lambda)=\sum_{k=1}^NE_{kk}-
h\lambda^{-1}\sum_{k=1}^N\gamma_kE_{k,k+1},
\end{equation}
\begin{equation}
{\cal D}(c,d,\lambda)=\sum_{k=1}^NE_{kk}-
h\lambda^{-1}\sum_{k=1}^N\delta_kE_{k,k+1},
\end{equation}
where $\gamma_k$'s are defined by (4.35), and $\delta_k$'s by (5.31) or (5.32)
below.}

{\bf Proof.} It is staightforward to check that the matrix equations (5.25) are
equivalent to the following ones:
\begin{equation}
\gamma_{k-1}\widetilde{d}_k=d_{k-1}\delta_{k-1},\quad
\widetilde{d}_k-h\gamma_k=d_k-h\delta_{k-1},
\end{equation}
\begin{equation}
\gamma_{k-1}\widetilde{c}_k=c_{k-1}\delta_k,\quad
\widetilde{c}_k+h\gamma_k=c_k+h\delta_k.
\end{equation}

Now it could be checked by means of direct calculation that the equations
(5.29), (5.30) are satisfied
with the following identifications: (4.32), (4.33) for $d_k$,
$\widetilde{d}_k$; $c_k=g^2
\exp(x_{k+1}-x_k)d_k$ and similarly for $\widetilde{c}_k$; $\delta_k=g^2
\exp(\wx_{k+1}-x_k)$, and
\[
\gamma_k=g^2\exp(\wx_{k+1}-x_k)
\frac{(\exp(-\wx_{k+1}+x_{k+1})-1)}{(\exp(-\wx_k+x_k)-1)}\:
\frac{(1+g^2\exp(x_k-x_{k-1}))}{(1+g^2\exp(x_{k+1}-x_k))},
\].

Working, however, again directly with (5.29), (5.30), without turning to the
representation through the $x_k$ variables, we note that (5.29) is equivalent
to
\begin{equation}
\widetilde{d}_k=d_{k-1}\frac{d_k+h\gamma_k}{d_{k-1}+h\gamma_{k-1}},\quad
\delta_{k-1}=\gamma_{k-1}\frac{d_k+h\gamma_k}{d_{k-1}+h\gamma_{k-1}},
\end{equation}
and (5.30) is equivalent to
\begin{equation}
\widetilde{c}_k=c_{k-1}\frac{c_k-h\gamma_k}{c_{k-1}-h\gamma_{k-1}},\quad
\delta_k=\gamma_{k-1}\frac{c_k-h\gamma_k}{c_{k-1}-h\gamma_{k-1}}.
\end{equation}
The first equations in (5.31), (5.32) coincide with (4.36), and the
compatibility of the second ones is equivalent to
\[
\frac{\gamma_k(d_{k+1}+h\gamma_{k+1})}{c_k-h\gamma_k}=
\frac{\gamma_{k-1}(d_k+h\gamma_k)}{c_{k-1}-h\gamma_{k-1}},
\]
which is a direct consequence of (4.35). The proof is complete.

Note that the matrices (5.27), (5.28) when compared with
(2.14), (2.15) satisfy
\[
{\cal C}(c,d,\lambda)=I+hC(c,d,\lambda)+O(h^2),\quad
{\cal D}(c,d,\lambda)=I+hD(c,d,\lambda)+O(h^2),
\]
as it follows from (4.37).

\setcounter{equation}{0}
\section{Factorization problems
and interpolating Hamiltonians}

It is very remarkable that the matrices ${\rm\bf A, B, C, D}$ and
${\cal A, B, C, D}$ from the previous
section may be identified with the certain factors $\Pi_{\pm}(f(T_{\pm}))$,
as in the Theorem 2.

{\bf Theorem 7.} {\it There hold following relations:}
\begin{eqnarray}
{\rm\bf A}(c,d,\lambda) & = & \Pi_+\left(I+hT_+(c,d,\lambda)\right),\\
{\rm\bf B}(c,d,\lambda) & = & \Pi_+\left(I+hT_-(c,d,\lambda)\right),\\
{\rm\bf C}(c,d,\lambda) & = &
\Pi_-^{-1}\left(I-hT_+^{-1}(c,d,\lambda)\right),\\
{\rm\bf D}(c,d,\lambda) & = & \Pi_-^{-1}\left(I-hT_-^{-1}(c,d,\lambda)\right).
\end{eqnarray}

{\bf Proof.} Define following two matrices:
\begin{eqnarray*}
Q_-(c,d,\lambda) & = & \sum_{k=1}^NE_{kk}-
\lambda^{-1}\sum_{k=1}^N\frac{c_k}{{\goth a}_k}E_{k,k+1} \in {\rm\bf G}_-,\\
Q_+(c,d,\lambda) & = & \sum_{k=1}^N\frac{c_k}{{\goth d}_k}E_{kk}+
\lambda\sum_{k=1}^NE_{k+1,k} \in {\rm\bf G}_+.
\end{eqnarray*}

Note now that the recurrent relation (4.14) is just equivalent to the matrix
equality
\begin{equation}
U(c,d,\lambda)+hL(c,d,\lambda)={\rm\bf A}(c,d,\lambda)Q_-(c,d,\lambda),
\end{equation}
and the recurrent relation (4.21) is just equivalent to the matrix equality
\begin{equation}
L(c,d,\lambda)-hU(c,d,\lambda)=Q_+(c,d,\lambda){\rm\bf D}(c,d,\lambda).
\end{equation}

Multiplying (6.5) from the right by $U^{-1}$, we obtain:
\[
I+hT_+={\rm\bf A}Q_-U^{-1},
\]
and since $Q_-U^{-1}\in {\rm\bf G}_-$, we obtain (6.1). From the previous
equation with the help of (5.1) we derive also
\[
I+hT_-=U^{-1}{\rm\bf A}Q_-={\rm\bf B}\widetilde{U}^{-1}Q_-,
\]
which proves (6.2) in view of $\widetilde{U}^{-1}Q_-\in {\rm\bf G}_-$.

Next, multiplying (6.6) from the left by $L^{-1}$, we obtain:
\[
I-hT_-^{-1}=L^{-1}Q_+{\rm\bf D},
\]
which just implies (6.4) because of $L^{-1}Q_+\in {\rm\bf G}_+$. Finally,
from the previous equation we derive with the help of (5.9):
\[
I-hT_+^{-1}=Q_+{\rm\bf D}L^{-1}=Q_+\widetilde{L}^{-1}{\rm\bf C},
\]
which means the validity of (6.3) because of $Q_+\widetilde{L}^{-1}\in
{\rm\bf G}_+$.

The theorem is proved.

{\bf Theorem 8.} {\it There hold following relations:}
\begin{eqnarray}
{\cal A}(c,d,\lambda) & = &
\Pi_+^{-1}\left((I-hT_+(c,d,\lambda))^{-1}\right),\\
{\cal B}(c,d,\lambda) & = &
\Pi_+^{-1}\left((I-hT_-(c,d,\lambda))^{-1}\right),\\
{\cal C}(c,d,\lambda) & = &
\Pi_-\left((I+hT_+^{-1}(c,d,\lambda))^{-1}\right),\\
{\cal D}(c,d,\lambda) & = & \Pi_-\left((I+hT_-^{-1}(c,d,\lambda))^{-1}\right).
\end{eqnarray}

{\bf Proof.} Define following two matrices:
\begin{eqnarray*}
P_-(c,d,\lambda) & = & \sum_{k=1}^NE_{kk}-
\lambda^{-1}\sum_{k=1}^N\frac{c_k}{\beta_{k+1}}E_{k,k+1} \in {\rm\bf G}_-,\\
P_+(c,d,\lambda) & = & \sum_{k=1}^N\frac{c_{k-1}}{\gamma_{k-1}}E_{kk}+
\lambda\sum_{k=1}^NE_{k+1,k} \in {\rm\bf G}_+.
\end{eqnarray*}

Note now that the recurrent relation (4.28) is just equivalent to the matrix
equality
\begin{equation}
U(c,d,\lambda)-hL(c,d,\lambda)=P_-(c,d,\lambda){\cal B}(c,d,\lambda),
\end{equation}
and the recurrent relation (4.35) is just equivalent to the matrix equality
\begin{equation}
L(c,d,\lambda)+hU(c,d,\lambda)={\cal C}(c,d,\lambda)P_+(c,d,\lambda).
\end{equation}

Multiplying (6.11) from the left by $U^{-1}$, we obtain:
\[
I-hT_-=U^{-1}P_-{\cal B},
\]
and since $U^{-1}P_-\in {\rm\bf G}_-$, we obtain (6.8). From the previous
equation with the help of (5.17) we derive also
\[
I-hT_+=P_-{\cal B}U^{-1}=P_-\widetilde{U}^{-1}{\cal A},
\]
which proves (6.7) in view of $P_-\widetilde{U}^{-1}\in {\rm\bf G}_-$.

Next, multiplying (6.12) from the right by $L^{-1}$, we obtain:
\[
I+hT_+^{-1}={\cal C}P_+L^{-1},
\]
which just implies (6.9) because of $P_+L^{-1}\in {\rm\bf G}_+$. Finally,
from the previous equation we derive with the help of (5.25):
\[
I+hT_-^{-1}=L^{-1}{\cal C}P_+={\cal D}\widetilde{L}^{-1}P_+,
\]
which means the validity of (6.10) because of $\widetilde{L}^{-1}P_+\in
{\rm\bf G}_+$.

The theorem is proved.

Substitute now the expressions (6.1)--(6.4) into the discrete Lax equations
(5.1), (5.2), (5.9), (5.10), and (6.7)--(6.10) into (5.17), (5.18), (5.25),
(5.26). One recognizes immediately the difference
equations from the part c) of the Theorem 2. This gives us the solution
of the initial v
alue problem for the dynamical system
(4.14), (4.15) in terms of the factorization of the matrices
\[
\left((I+hT_{\pm}(0))\right)^n,
\]
the solution of the initial value problem for the dynamical system
(4.21), (4.22) in terms of the factorization of the matrices
\[
\left((I-hT_{\pm}^{-1}(0))\right)^n.
\],
the solution of the initial value problem for the dynamical system
(4.28), (4.29) in terms of the factorization of the matrices
\[
\left((I-hT_{\pm}(0))^{-1}\right)^n.
\],
and the solution of the initial value problem for the dynamical system
(4.35), (4.36) in terms of the factorization of the matrices
\[
\left((I+hT_{\pm}^{-1}(0))^{-1}\right)^n.
\],

The part d) of the Theorem 2 may be formulated in our case in the following
lines.

{\bf Corollary.} {\it The interpolating Hamiltonian for the map
((4.14), (4.15) is given by
\[
\varphi_+(T)={\rm tr}(\Phi_+(T))=J_+(T)+O(h), \quad {\rm where}\quad
\Phi_+(\xi)=h^{-1}\int_0^{\xi}\frac{d\eta}{\eta}\log(1+h\eta),
\]
for the map  (4.21), (4.22) -- by
\[
\varphi_-(T)={\rm tr}(\Phi_-(T))=J_-(T)+O(h), \quad {\rm where}\quad
\Phi_-(\xi)=h^{-1}\int_{\xi}^{\infty}\frac{d\eta}{\eta}
\log(\frac{1}{1-h\eta^{-1}}),
\]
 for the map  (4.28), (4.29) -- by
\[
\psi_+(T)={\rm tr}(\Psi_+(T))=J_+(T)+O(h), \quad {\rm where}\quad
\Psi_+(\xi)=h^{-1}\int_0^{\xi}\frac{d\eta}{\eta}\log\frac{1}{1-h\eta},
\]
and for the map (4.35), (4.36) -- by
\[
\psi_-(T)={\rm tr}(\Psi_-(T))=J_-(T)+O(h), \quad {\rm where}\quad
\Psi_-(\xi)=h^{-1}\int_{\xi}^{\infty}\frac{d\eta}{\eta}
\log(1+h\eta^{-1}).
\]
}

\section{Conclusion}
We have introduced the difference approximations for two Hamiltonian
flows from the relativistic Toda hierarchy. They turned out to belong
to the same hierarchy. The inclusion in the general scheme of symplectic
maps on groups equiped with quadratic Poisson brackets allowed to solve
the difference equations in terms of factorization problem in the group
and to  find the interpolating Hamiltonians.

\newpage
\section{Acknowledgements}
The research of the author is financially supported by the Deutsche
Forschungsgemeinschaft.

The substantial part of this work was done during my one--week visit
to the University of Rome III. I cordially thank Professor Orlando
Ragnisco for hosting this visit, as well as for the most helpful
encouragement and collaboration.

\end{document}